\documentclass[12pt, a4paper, oneside]{article}
\usepackage[utf8]{inputenc}
\usepackage{amsmath}
\usepackage{amssymb}
\usepackage{mathtools}
\usepackage[english]{babel}
\usepackage{epsfig}
\usepackage[usenames,dvipsnames]{xcolor}
\usepackage{float}
\usepackage{setspace}
\usepackage{natbib}
\usepackage{sectsty}
\newcounter{ourcount}
\usepackage{authblk}

\subsectionfont{\normalfont\textit}
\subsubsectionfont{\normalfont\textit}

\title{On Evaluation of Risky Investment Projects. Investment Certainty Equivalence}

\author[a,b]{Andrey Leonidov\footnote{leonidovav@lebedev.ru}}

\author[a]{Ilya Tipunin\footnote{tipunin@gmail.com}}

\author[a,b]{Ekaterina Serebryannikova\footnote{serebryannikovaee@lebedev.ru}}

\affil[a]{\footnotesize P.N. Lebedev Physical Institute, 53, Leninsky prospect, Moscow, Russia, 119333}
\affil[b]{\footnotesize Moscow Institute of Physics and Technology, 9 Institutskiy per., Dolgoprudny, Moscow Region, Russia, 141701}

\begin{document}

\maketitle

	\begin{abstract}
	The purpose of the study is to propose a methodology for evaluation and ranking of risky investment projects.
	An investment certainty equivalence approach dual to the conventional separation of riskless and risky contributions based on cash flow certainty equivalence is introduced. Proposed ranking of investment projects is based on gauging them with the Omega measure, which is defined as the ratio of chances to obtain profit/return greater than some critical (minimal acceptable) profitability over the chances to obtain the profit/return less than the critical one.
	Detailed consideration of alternative riskless investment is presented. Various performance measures characterizing investment projects with a special focus on the role of reinvestment are discussed. Relation between the proposed methodology and the conventional approach based on utilization of risk-adjusted discount rate (RADR) is discussed. Findings are supported with an illustrative example.
	 The methodology proposed can be used to rank projects of different nature, scale and lifespan. In contrast to the conventional RADR approach for investment project evaluation, in the proposed method a risk profile of a specific project is explicitly analyzed in terms of appropriate performance measure distribution. No ad-hoc assumption about suitable risk-premium is made.

\end{abstract}

\section{Introduction}

	Evaluation and ranking of investment projects is one of the most important problems in corporate finance \citep{BMA}. At the heart of this problem is a necessity of formulating investment goal taking into account the variability of the project outcomes. 
	
	A risky investment project is naturally defined as a project that could bring return exceeding that of the alternative riskless investment generating the  same cash flow pattern. A risk premium corresponds to return above the riskless one that, however, is not guaranteed, i.e. is uncertain. Therefore the possible outcomes for risk premium are specified in terms of a probability distribution that provides a quantitative description of its variability.  A decision to go for a project is thus contingent on the estimate of chances of achieving investor's profit benchmarks that depend on the shape of the risk premium probability distribution.  To our knowledge, this idea was first formulated in \citep{Fishburn}.  
	
	Let us note that historically the mean return and its standard deviation were used as project "coordinates" \,in the risk-return space \citep{Markowitz,BP}. This works fine for symmetric return distributions which shape is close to that of the Gaussian (normal) distribution. However, it was soon realized that a much better characterization of risk-return space is achieved by replacing mean return and its standard deviation  by the required return and some asymmetric risk measure taking into account that the notion of risk is naturally related only to returns smaller than the required one  \citep{BP,Caporin,Krochmal,Laughunn,Miller,Nawrocki,Sortino}. For recent applications of this method to evaluation of complex investment projects see e.g. \citep{DMR07,Leite}.
	
Assessment of the quality of the project under consideration proceeds through two basic steps:
\begin{itemize}
\item specification of the acceptable risky profit range, in particular - of the lowest acceptable profit in terms of a risk premium defined with respect to an appropriate riskless benchmark,
\item quantification of chances for the risky profit to miss the acceptable range (risk) as following from the analysis of variability of risk premium characterized by the shape of its probability distribution.
\end{itemize}
A project is thus  characterized by a given (chosen by the investor) hurdle risk premium and uncertainty associated with chances of its realization. This also lays a basis for ranking several projects: this is done by ranging, for a given threshold risk premium, values of some suitably defined measure of missing the target risk premium.

A widely used approach to this problem is to use averaged cash flows in combination with a risk-adjusted discount rate (RADR) \citep{BMA}. It is, however, known that the RADR
approach is not universal and is applicable only to investment projects possessing specific characteristics \citep{RM66,MT77,F77}, see also \citep{H86}. 
In particular, in \citep{RM66} it was shown that using RADR one implicitly fixes very special structure of the investor's preferences. This follows from comparison of the RADR estimate and the expression obtained by using the certainty equivalence approach. An interesting recent development along these lines was described in \citep{EM13,E14}. 	

In the present paper we propose a method of separation of riskless and risky contributions of investment project cash flows based on constructing a replicating riskless investment for each possible cash flow realization. We term the corresponding principle Investment Certainty Equivalence. This is, in particular, to stress that an important question of reinvestment should be treated separately, see e.g. a recent discussion in \citep{C17}. With respect to interrelation between the RADR and certainty equivalence approaches it is shown that fixing some risk-adjusted discount rate implies a certain separation of  riskless and risky contributions to the cash flows. As to the quantitative assessment  of risk/return profile we will use  a particularly interesting asymmetric risk/return measure, the so-called $\Omega$, considered in \citep{KSG,SK,Bertrand}.

The main objective of the paper is to describe a new methodology of investment projects ranking. Therefore detailed description of  sources of risks and of methodologies of their modelling as well as a detailed analysis of mechanisms underlying projects cash flows are out of the scope of the current study. However, some comments on classification of risk factors can be made. All the risks that influence efficiency of investment projects can be divided into two classes:
cash flow risks and alternative investment/reinvestment risks. The first class contains such risk factors as variability of macroeconomic indicators, market prices or operational risks. Such risks exert direct influence on  project's cash flows. The second class contains risks related to variability of reinvestment rate. As this paper analyses riskless alternative only, the methodology of accounting for risks of the second class is out of the scope of the present analysis and consitutes an interesting direction for future studies.

The paper is organized as follows. 

In the Section \ref{ERRP} we detail a procedure of evaluation and ranking of risky projects. In paragraph \ref{IPGD} we outline a general description of the key ingredients of an investment project as well as 
quantitative characteristics used in its assessment. In paragraph \ref{EIPO} we give a schematic outline of the evaluation procedure considered in the the present paper. In paragraph \ref{IPARI} we construct a riskless portfolio replicating a given cash flow stream. In paragraph \ref{IPC} we describe key quantitative characteristics used in evaluation of an investment project. In paragraph \ref{IPRPM} we discuss risk premium measures that arise in the approach under consideration. In paragraph \ref{IPDBRP} we describe a quantitative criterion suggested to make an investment-related decision based on the risk profile of an investment project. In paragraph \ref{critval} we derive formulae for the critical/threshold values for the key quantitative characteristics of an investment project. 

In the Section \ref{RADR} we discuss some quantitative aspects of comparison between the conventional RADR approach and the approach discussed in the present paper. In paragraph \ref{RADRMO} we outline the conventional RADR methodology. In paragraph \ref{RADRE} we discuss a conventional certain equivalence approach. In paragraph \ref{RADRD} we discuss the relation between the RADR approach and the one suggested in the present paper. 

In the Section \ref{example} we compare the results of ranking of two model investment projects using RADR and suggested approaches correspondingly.

In the Section \ref{real_ex} the example of ranking of real industrial projects is provided.

In the Section \ref{conclusion} we present our conclusions.

\section{Evaluation and ranking of risky projects}\label{ERRP}
	
\subsection{Investment project: general description}\label{IPGD}

In this paragraph we focus on the description of projects with the simplest structure of cash flows with the single negative contribution corresponding to an initial investment. The generalization for the case of arbitrary structure of cash flows  is described in Paragraph~\ref{IPARI}. 

In what follows we assume that a description of an investment project of duration $T$ to be assessed is available in the following form. A project involves two major interrelated processes, those of investment and reinvestment:
\begin{itemize}
\item {\bf Investment} \\
\\
 Investment process corresponds to a transformation of an initial investment outlay $I_0$ into one of the possible realizations $\{ {\bf F}^{(i)} \}$ of the cash flow stream generated by the project 
\begin{equation}\label{ipdfull}
I_0 \longmapsto 
\begin{cases}
{\bf F}^{(1)}=(F^{(1)}_1\;\; , \cdots , & F^{(1)}_T )\\
\;\;\;\;\;\;\;\;\; , \cdots , &  \\
\medskip
{\bf F}^{(N)}=(F^{(N)}_1\;\; , \cdots , & F^{(N)}_T),
\end{cases}
\end{equation}
or, in condensed notation,
\begin{equation}
I_0 \; \underset{ \text{investment}}{\longmapsto} \; \{ {\mathbf F}^{(i)} \}, \;\;\;\; i = 1, \cdots, N,
\end{equation}
where cash flow streams ${\bf F}^{(i)}=(F^{(i)}_1, \cdots , F^{(i)}_T )$ correspond to $N$ possible outcomes that are, e.g., generated through Monte-Carlo simulation or scenario analysis. The uncertain nature of the project outcome makes it natural to use a probabilistic description where, generically, a project is fully characterized by some multinomial probability distribution ${\cal P}_F(F_1, \cdots, F_T)$:
\begin{equation}\label{ipdprob}
I_0  \underset{ \text{investment}}{\longmapsto} \;  {\cal P}_F (F_1, \cdots, F_T)
\end{equation}
\item {\bf Reinvestment} \\
\\
Reinvestment process corresponds to transformation of cash flow streams $\{ {\bf F}^{(i)} \}$ into the set of terminal cash flows at maturity T $\{ F^{{\rm tot} (i)}_T \}$ through reinvesting\footnote{In general the partial cash flows $F_t$ can be both positive (profits) and negative (additional investments). Of course, reinvestment process operates only with  intermediate positive cash flows.} the components of $\{ {\bf F}^{(i)} \}$ into the same project or some other riskless/risky projects (reinvestment) providing (uncertain) forward profitability rates ${\bf K}^f = (K^f_{1 \to T}, K^f_{2 \to T}, \cdots, K^f_{T-1 \to T})$ described by a probability distribution ${\cal P}_{K^f} ({\bf K}^f)$ such that for each intermediate cash flow we have
\begin{equation}
F^{(i)}_t (1+K^f_{t \to T}) = F^{{\text tot} (i)}_T \equiv {\rm FV} (F^{(i)}_t \vert K^f_{t \to T}),
\end{equation}
where ${\rm FV} (F^{(i)}_t \vert K^f_{t \to T})$ denotes the future value at time $T$ of the partial cash flow $F^{(i)}_t $ corresponding to the partial rate $K^f_{t \to T}$.  Therefore, for each realization of the cash flow stream we have
 \begin{equation}
 \{ {\mathbf F}^{(i)} \}   \; \underset{ \text{reinvestment}}{\overset{{\bf K}^f}\longmapsto}  \; \{ {\rm FV} (F^{(i)}_t \vert K^f_{t \to T}) \}
 \end{equation}
 or, in condensed notation,
\begin{equation}
{\bf F}  \; \underset{ \text{reinvestment}}{\overset{{\bf K}^f}\longmapsto}   {\rm FV} ({\bf F} \vert {\bf K}^f)
\end{equation}
 
 Within probabilistic description the reinvestment process generates a final probability distribution of terminal cash flow ${\cal P}_{F^{\rm tot}_T} (F_T^{\text tot})$:
\begin{equation}
{\cal P}_F(F_1, \cdots, F_T) \; \underset{ \text{reinvestment}}{\overset{{\cal P}_{K^f}({\bf K}^f)}\longmapsto} \; {\cal P} _{F^{\rm tot}_T} (F^{\text tot}_T)
\end{equation}
More explicitly,
\begin{equation}
{\cal P}_{F^{\rm tot}_T}  (F^{\text tot}_T) = \int  d {\bf K}^f d {\bf F} \; {\cal P}_{K^f} ({\bf K}^f)  {\cal P}_F ({\bf F}) \; \delta \left( F^{\text tot}_T - {\rm FV} ({\bf F} \vert {\bf K}^f) \right)
\end{equation} 
\end{itemize}

The full description of an investment project can therefore be summarized by the following superposition of investment and reinvestment processes:
\begin{equation}\label{InvProc}
I_0 \; \underset{ \text{investment}}{\longmapsto} \; \{ {\mathbf F}^{(i)} \} 
\; \underset{ \text{reinvestment}}{\overset{{\bf K}^f}\longmapsto}  
\; \{ F^{(i){\text tot }}_T = {\rm FV} ({\bf F}^{(i)} \vert {\bf K}^f) \}
\end{equation}
or, in probabilistic terms, as
\begin{equation}\label{InvProc_prob}
I_0  \; \underset{ \text{investment}}{\longmapsto} \;  {\cal P}_F ({\bf F})   \; \underset{ \text{reinvestment}}{\overset{{\cal P}({\bf K}^f)}\longmapsto} \; {\cal P}_{F^{\rm tot}_T}  (F^{\text tot}_T= {\rm FV} ({\bf F} \vert {\bf K}^f))
\end{equation}

From the  distribution of terminal cash flow ${\cal P}_{F^{\rm tot}_T}  (F_T^{\text tot})$  one can calculate the distributions of the project profit 
$ \Pi_T=F_T^{\text tot}-I_0$
\begin{equation}\label{disP}
{\cal P}_{\Pi} (\Pi_T) = \int d F^{\rm tot}_T \delta ( \Pi_T -F_T^{\text tot}+I_0) {\cal P}_{F^{\rm tot}_T}  (F_T^{\text tot}) =  
{\cal P}_{F^{\rm tot}_T}  (\Pi_T+I_0)
\end{equation}
 and/or its return $ M_T=(F_T^{\text tot}-I_0)/I_0$ (or, in the annualized form, $M_T \equiv (1+\mu)^T-1$)
\begin{equation}\label{disM}
{\cal P}_M (M_T) =  \int d F^{\rm tot}_T \delta ( M_T -\frac{F_T^{\text tot}-I_0}{I_0}) {\cal P}_{F^{\rm tot}_T}  (F_T^{\text tot}) =
I_0 {\cal P}_{F^{\rm tot}_T}  (I_0 (M_T-1)).
\end{equation}

Although some details of the full probabilistic description like smoothness of the intermediate incomes can be of interest for evaluating the quality of an investment project, the consideration is usually restricted to analyzing the properties of the distributions ${\cal P}_{\Pi} (\Pi_T)$ and/or  ${\cal P}_M (M_T)$ that, as described above, are fully determined, see \eqref{disP} and \eqref{disM}, by the distribution of the terminal cash flow ${\cal P}_{F^{\rm tot}_T}  (F_T^{\text tot})$.

\subsection{Evaluation of an investment project: an outline}\label{EIPO}

In this section we provide a logical outline of the proposed method of investment evaluation against riskless alternative and define a notion of investment certainty equivalence. For convenience we break the procedure into stages as follows: 
	
\begin{enumerate}
\item As described  in the previous paragraph, a risky investment project with duration $T$ can generically be described as a superposition of two processes:
\begin{itemize}
 \item  transformation of initial investment $I_0$ into a cash flow stream ${\bf F}=(F_1, \cdots, F_T)$ realizing possible outcomes of investing $I_0$ into the particular project under consideration (investment) 
 \begin{equation}
 I_0 \; \underset{ \text{investment}}{\longmapsto} \; {\mathbf F}; 
 \end{equation}  
 \item transformation of  ${\bf F}$ into a terminal cash flow at the end of the project $F^{\rm tot}_T$ through reinvesting ${\bf F}$ into the same or some other riskless or risky projects (reinvestment):
 \begin{equation}
 {\mathbf F}   \; \underset{ \text{reinvestment}}{\longmapsto} \; F^{\rm tot}_T.
 \end{equation}
 \end{itemize}
\item An {\it investment certainty equivalent} is defined as a riskless investment ${\tilde I}_0$ generating the same cash flow pattern ${\bf F}=(F_1, \cdots, F_T)$:
\begin{equation}
 {\tilde I}_0 \; \underset{ \text{riskless investment}}{\longmapsto} \; {\mathbf F}.
\end{equation}
The difference ${\tilde I}_0-I_0$ between the investment certainty equivalent and the project investment outlay quantifies the investment risk premium. The notion of investment certainty equivalence is dual to the conventional certainty equivalence related to the riskless investment of the original investment outlay  $I_0$ producing a modified cash flow pattern, see discussion in the paragraph \ref{RADR} below.
\item Generically the outcomes of both investment and reinvestment are uncertain and, therefore, both processes are risky. A risk premium (a gap between risky and riskless return/profit) does thus include two different contributions so that generically there exist two different risk premiums corresponding to uncertainties in investment and reinvestment.
\item The present study is mainly focused at investment risk premium and assumes riskless alternative investment and riskless reinvestment.\footnote{For a discussion of some features of risky alternative investment  see paragraph \ref{RADR}.}
\item For investment project to be attractive it should with acceptable probability bring profit exceeding the minimally acceptable one specified by the investor \citep{Fishburn}. 
\item The risk related to risk premiums is quantified by analyzing their distributions and evaluating the corresponding quantities characterizing the risk/return profile of the investment project. In what follows we shall restrict our consideration to the  parameter $\Omega$ (defined below in 
Paragraph~\ref{IPDBRP} expression
\eqref{omega}) characterizing  risk/return relation. 
\item The projects are then accepted/ranged according to the investor risk/return preferences. Namely, the projects are ranked in decreasing order in $\Omega$, the projects with highest values of $\Omega$ being the best. 
\end{enumerate}
	
\subsection{Investment project: alternative riskless investment}\label{IPARI}

The suggested method of evaluation and ranking of investment projects  is based on gauging  investment $I_0$ generating the set of expected cash flow trajectories against the {\it riskless} alternatives generating the same set of cash flow trajectories and characterized by the riskless yield curve ${\bf R}=(R_1, \cdots,R_T)$ or, equivalently, its annualized counterpart ${\bf r}=(r_1, \cdots,r_T)$ and the riskless reinvestment forward rates ${\bf R}^f=(R_1^f, \cdots,R^f_T)$ and ${\bf r^f}=(r_1^f, \cdots,r^f_T)$ which can be calculated from the riskless yield curve. The exact procedure is described below. 

	In the case of riskless 
	reinvestment the expression ~\eqref{InvProc}  describing investment and reinvestment processes takes the following form:
\begin{equation}\label{InvProc_riskless}
I_0 \; \underset{ \text{investment}}{\longmapsto} \; \{ {\mathbf F}^{(i)} \} 
\; \underset{ \text{reinvestment}}{\overset{{\bf R}^f}\longmapsto}  
\; \{ F^{(i){\text tot }}_T = {\rm FV} ({\bf F}^{(i)} \vert {\bf R}^f) \}
\end{equation}
or, in probabilistic terms (cf. equation \eqref{InvProc_prob}),
\begin{equation}\label{InvProc_prob_riskless}
I_0  \; \underset{ \text{investment}}{\longmapsto} \;  {\cal P}_F ({\bf F})   \; \underset{ \text{reinvestment}}{\overset{{\bf R}^f}\longmapsto} \; {\cal P}_{F^{\rm tot}_T}  (F^{\text tot}_T= {\rm FV} ({\bf F} \vert {\bf R}^f))
\end{equation}

In general a set of cash flows ${\mathbf F}$ contains both positive  ${\mathbf F}^+$ and negative ${\mathbf F}^-$ contributions corresponding to profits and additional future investment outlays correspondingly:
\begin{equation}
{\mathbf F}  = {\mathbf F}^+ - {\mathbf F}^-,
\end{equation}
where  
\begin{equation}
F^+_t = \max(F_t,0), \;\;\;\; F^-_t = \max(-F_t,0)
\end{equation}
Constructing a proper treatment of negative contributions to cash flow having natural interpretation of future additional investments is a subtle issue \footnote{For an early  discussion see e.g. \citep{B78,B82,MC79}.}. In the considered case of riskless alternative investment/reinvestment universe  the procedure is, however straightforward. The future investments can be guaranteed by an additional initial investment outlay 
\begin{equation}\label{PVFM}
{\rm PV} ({\mathbf F}^- \vert {\bf R}) = \sum_{t=1}^T \frac{F^-_t}{1+R_t}
\end{equation}
Such an additional investment can be arranged by buying  a portfolio of couponless riskless bonds with payments replicating future investments $F^-_t$ at the appropriate time horizons. The portfolio consists from partial investments $(I_0^{(1)}, \cdots, I_0^{(T)})$ such that 
\begin{equation}
I_0^{(t)} (1+R_t)=F^-_t \;\; \to \;\; \sum_t I_0^{(t)}=   \sum_{t=1}^T \frac{F^-_t}{1+R_t}
\end{equation}
so that for the particular cash flow pattern under consideration the initial investment outlay should include the additional investment \eqref{PVFM}, and, therefore, the investment pattern in~\eqref{InvProc_riskless} is for this realization replaced by
\begin{equation}\label{inv1}
I_0^{\text tot}\equiv I_0 + {\rm PV}({\bf F}^-|{\bf R})\; \underset{ \text{investment}}{\longmapsto} \; {\mathbf F}^{+} 
\end{equation}

Let us now turn to an explicit description of the riskless reinvestment pattern and note that the cash flow pattern ${\mathbf F}^+$ can be arranged by the riskless investment ${\tilde I}_0$ through investing into  a portfolio of bonds $(B_1, \cdots, B_T)$ such that
\begin{equation}
B_t (1+R_t) = F^+_t, \;\;\;\; {\tilde I}_0=\sum_{t=1}^T B_t = {\rm PV}( {{\mathbf F}^+ \vert {\mathbf R}} )
\end{equation} 
and assume that at $t=0$ we fix 
a forward contract for buying  at time $t$ at the price $F^+_t$  the bond maturing at $T$ thus fixing the corresponding rate $R^f_{T-t}$. This leads to the cash flow at $T$ equalling $F^+_t(1+R^f_{T-t})=B_t(1+r_t)^t(1+R^f_{T-t})$. On the other hand the same cash flow can be fixed by buying at $t=0$ the bond maturing at $T$ so that  
$B(1+r_T)^T=B_t(1+r_t)^t(1+R^f_{T-t})$. The two riskless portfolios giving the same profit should have equal initial investments at $t=0$, i.e.  $B_t=B$. We obtain therefore 
\begin{equation}
(1+r_T)^T=(1+r_t)^t(1+R^f_{T-t}),
\end{equation} 
thus fixing the forward rate in question
\begin{equation}
\label{forvard}
R^f_{T-t}=\frac{(1+r_T)^T}{(1+r_t)^t}-1.
\end{equation}
The vector of positive cash flows ${\bf F^+}$ has the same riskless present value as the riskless cash flow $\sum_{t=1}^{T}F^+_t(1+R^f_{T-t})$ and thus we have indeed fixed the forward rate curve determining the forward value
\begin{equation}
{\rm FV} ({\mathbf F}^+ \vert {\bf R}^f) = \sum_{i=1}^T F^+_i (1+R^f_{T-i}), 
\end{equation}
so that the complete description of some particular outcome of an investment project taking into account the necessity of additional investment outlays can be described as
\begin{equation}\label{invreinv2}
I_0^{\text tot}\equiv I_0 + {\rm PV}({\bf F}^-|{\bf R})\; \underset{ \text{investment}}{\longmapsto} \; {\mathbf F}^{+} 
\; \underset{ \text{reinvestment}}{\overset{{\bf R}^f}\longmapsto}  
\; F^{\text tot}_T = {\rm FV} ({\bf F}^{+} \vert {\bf R}^f)
\end{equation}

\subsection{Investment project: characteristics}\label{IPC}

The profitability of an investment project on each cash flow trajectory can be characterized in several ways. The list of the corresponding characteristics includes, in particular,
\begin{itemize}
\item the net terminal profit $\Pi_T ({\mathbf F})$;
\begin{equation}
\Pi_T= {\rm FV} ({\mathbf F}^+\vert {\mathbf R}^f) - I_0 - {\rm PV} ({\mathbf F}^- \vert {\mathbf R}) \equiv {\rm FV} ({\mathbf F}^+\vert {\mathbf R}^f) - I_0^{\text{tot}} 
\end{equation}
\item the terminal return $M_T$ and its annualized version $\mu$
\footnote{The meaning of $\mu$ if close to that of ${\rm MIRR}(k,d)$ defined by
$$
(1+{\rm MIRR}(k,d))^T=\frac{\sum_{t=1}^TF_t^+(1+k)^{T-t}}{\sum_{t=0}^TF_t^-(1+d)^{-t}}.
$$		
Let us stress  the above-defined MIRR assumes the flat term structure structure of both the reinvestment and financing rate curves (see also\citep{Kierulff2008}).
}
\begin{equation}
1+M_T = (1+\mu)^T= \frac{ {\rm FV} ({\mathbf F}^+\vert {\mathbf R}^f)}{I_0^{\text{tot}}};
\end{equation}
\item the net present value 
\begin{eqnarray}\label{defNPV}
{\rm NPV} ({\mathbf F}) &=& {\tilde I}_0 - \left(I_0 + {\rm PV} ({\mathbf F}^- \vert {\mathbf R}) \right)  \nonumber \\
& \equiv & {\rm PV} ({\mathbf F}^+\vert {\mathbf R}) - I_0^{\text{tot}} \equiv {\rm PV}({\mathbf F} \vert {\mathbf R}) - I_0
\end{eqnarray}
\item the profitability index
\begin{equation}
{\rm PI} = \frac{ {\rm NPV} ({\mathbf F}) }{I_0^{\text{tot}}}.
\end{equation}
\end{itemize}
Let us stress that evaluation of the risk/return profile of an investment project does depend on the target characteristics chosen by an investor.

\subsection{Investment project: risk premium measures}\label{IPRPM}

	An amount of risk premium collected by an investor on the particular trajectory described in \eqref{invreinv2} can be quantified by comparing the risky investment \eqref{invreinv2} with its riskless alternative 
$\tilde{I}_0$. In the case under consideration the two investments to compare are

\begin{eqnarray}
	I_0^{tot} \equiv I_0 + {\rm PV} ({\mathbf F}^- \vert {\mathbf R}) & \underset{ \text{investment}}{\longmapsto} &  {\mathbf F}^+ \label{inv3} \\
	\tilde{I}_0 \equiv {\rm PV} ({\mathbf F}^+ \vert {\mathbf R}) & \underset{ \text{investment}}{\overset{{\bf R}}\longmapsto} &  {\mathbf F}^+ \label{rfinv2}
\end{eqnarray}

The risky investment \eqref{inv3} is preferable to the riskless alternative \eqref{rfinv2}  if
\begin{equation}
I_0 + {\rm PV} ({\mathbf F}^- \vert {\mathbf R}) < {\rm PV} ({{\mathbf F}^+ \vert {\mathbf R}})
\end{equation}
i.e. (see \eqref{defNPV}) if 
\begin{equation}
{\rm NPV}({\mathbf F} \vert {\mathbf R}) >0
\end{equation}
The corresponding risk premium $\Delta_{\text{NPV}}$ is thus simply equal to ${\rm NPV}({\mathbf F} \vert {\mathbf R})$.

Let us now find an explicit expression for the risk premium $\Delta_M$ for the terminal return $M \equiv R_T + \Delta_M$. This follows directly from
\begin{eqnarray}
1+R_T + \Delta_M & = & \frac{{\rm FV} ({\mathbf F}^+ \vert {\mathbf R}^f)}{I_0 + {\rm PV} ({\mathbf F}^- \vert {\mathbf R})} \nonumber \\
1+R_T  & = & \frac{{\rm FV} ({\mathbf F}^+ \vert {\mathbf R}^f)}{{\rm PV} ({\mathbf F}^+ \vert {\mathbf R})}
\end{eqnarray}
so that
\begin{equation}
\Delta_M = (1+R_T) \frac{ \Delta_{\text{NPV}}}{I_0^{\text(tot)}} \equiv (1+R_T) {\rm PI} ({\mathbf F} \vert {\mathbf R})
\end{equation}

\subsection{Investment project: decision based on risk profile}\label{IPDBRP}
	
For the riskless investments the risk premium is, obviously, absent, $\Delta_{\text{NPV}}=\Delta_M=0$. 
For risky cash flows the contributions  $F_t, \ t\ge0$  are not certain and, as a result, the investment project is characterized by the probability distribution of risk premium  ${\cal P}( \Delta_{\rm NPV})$  or ${\cal P}(\Delta_M)$ corresponding to the set of possible cash flow trajectories.  The investor's evaluation of the project should take this into account. A natural way of dealing with the uncertainty of the risk premium is to fix a hurdle risk premium $\Delta_{\text{NPV}}$ or  $\Delta_M$ and quantify risk by analyzing the chances of  the project risk premium being below this target. A natural way of weighting risks against gains is to use the ratio $\Omega$ \citep{SK,KSG} with some threshold risk premium scale $\Lambda$  
separating desirable and undesirable risk premium outcomes:
	\begin{equation}\label{omega}
	\Omega=\frac{\int_{\Lambda}^{\infty}dx (x-\Lambda) {\cal
            P}(x)}{\int_{-\infty}^{\Lambda}dx (\Lambda-x) {\cal P}(x)} \equiv
        \frac{{\rm Call}(\Lambda)}{{\rm Put}(\Lambda)} 
	\end{equation}
The second equality in \eqref{omega} reflects the fact that the ratio $\Omega$ has, for the risk premium $\Delta_{\text{NPV}}$, a natural interpretation in terms of the ratio of prices of the so-called Bachelier \citep{BP} call and put options, see \citep{KSG}. Let us stress that these prices are different from the commonly considered Black-Scholes ones, see a detailed discussion in  \citep{BP}. 
In this case equation \eqref{omega} ${\rm Call} (\Lambda)$ is a price of an European option on buying the risk premium at a price $\Lambda$ while ${\rm Put}(\Lambda)$ is that of a European option on selling it for the same price. The final decision on the project does thus depend on whether the value of $\Omega$ associated with the required risk premium characteristics $\Delta_{\text{NPV}}$ or $\Delta_M$ is acceptable in terms of the investor's risk/return considerations. 

The key property of $\Omega$ is that it is a monotonously decreasing function of the threshold risk premium. This allows to establish a simple criterion for admissible risk limiting the corresponding risk premium:
\begin{equation}
\Lambda: \Omega(\Lambda) \geq 1
\end{equation}
Let us note that the choice $\Omega(\Lambda)=1$ corresponds to the threshold value of $\Lambda$ equal to the mean risk premium and, therefore, to the risk-neutral choice. In turn, the choices corresponding to $\Omega(\Lambda)>1$  and  $\Omega(\Lambda)<1$ reflect risk averse and risk seeking choices respectively.
	
\subsection{Investment project: threshold values}\label{critval}

In the general case the threshold ${\rm NPV}^*$ for ${\rm NPV}$(or equivalently to $\Delta_{\rm NPV}$) provides the scale separating the distribution in ${\cal P}({\rm NPV})$ (or ${\cal P}(\Delta_{\rm NPV})$) into domains corresponding to gains/losses. The value of ${\rm NPV}^*$ can be fixed in different ways. It can be fixed by management based on the minimal acceptable gain $\Pi^*$, this directly determines  ${\rm NPV}^*$. Alternatively, it can be calculated from the value of minimally acceptable profitability of the project. 	

Let us assume that one specifies the risk premium $\Delta^*_{\mu}$. This means that 
$\mu$ should exceed $r_T+\Delta^*_{\mu}$. Defining 
$\mu^*=r_T+\Delta^*_{\mu}$ we get, for a given cash flow trajectory, an acceptance criterion	
\begin{equation}
\mu > \mu^*.
\end{equation}
The value ${\rm NPV}^*$ of the threshold risky NPV  corresponding to 
$\mu^*$ should satisfy
\begin{equation}
\label{crit_val_PV1}
-I_0+ \text{PV} ({\mathbf F} \vert {\mathbf r})={\rm NPV^*}\ \Longleftrightarrow\ \mu=\mu^*.
\end{equation}
Based on ~\eqref{crit_val_PV1} 
we get (for details see Appendix)
\begin{eqnarray}
\mu^*& = & r_T+\Delta^*_{\mu},\\
\label{NPV*}
{\rm NPV}^* & = & \left(\frac{(1+r_T+\Delta^*_{\mu})^T}{(1+r_T)^T}-1\right)\left(I_0+\sum_{t=1}^{T} I_0^{(t)}\right).
\end{eqnarray}

\section{Investment evaluation using RADR}\label{RADR}

Let us now turn to the comparison with the widespread methodology of an investment project valuation  - the risk-adjusted discount rate formalism (RADR) and  apply the same approach as in previous paragraphs to its description. For simplicity in this paragraph we will consider only the case of the 
simplest canonical cash flow (i.e. only positive cash flows at $t \geq 1$) and assume the flat term structure of the riskless rate.

\subsection{RADR: methodology outline}\label{RADRMO}

The standard algorithm of valuation within RADR approach includes the following three stages:
\begin{enumerate}
	\item The ensemble of cash flow trajectories is characterized by the vector of averages $\langle\mathbf{F}\rangle=(\langle F_1\rangle,\dots,\langle F_T\rangle)$. In the case when the ensemble of $\mathbf{F}$ consists of  $N$ realisations $\{\mathbf{F}^{(i)}\}$ this vector is generated by "vertical"\; averaging:
	\begin{eqnarray}\label{vert_means}
	I_0 &\longmapsto& 
	\begin{cases}
	{\bf F}^{(1)}=(F^{(1)}_1\;\; , \cdots , F^{(1)}_T )\\
	\;\;\;\;\;\;\;\;\;  \cdots   \\
	{\bf F}^{(N)}=(F^{(N)}_1\;\; , \cdots ,  F^{(N)}_T)
	\end{cases}\\\nonumber
	& &  \;\;\;\;\;\;\;\;\;\;\; \Downarrow  \nonumber \\
	 & &\ \ \langle\mathbf{F}\rangle\ \ =\ (\langle F_1\rangle\;\; , \cdots , \langle F_T\rangle)
	\end{eqnarray}
	
	\item The mean cash flows $\langle F_t \rangle,\ t=1,\dots,T,$ are discounted with the effective (risk-adjusted) rate $k=r+\Delta_r$.
	
	\item The initial investments $I_0$ are compared with the sum of discounted values of $\langle F_t\rangle$. The investor goes into a project if  
	\begin{equation}
	I_0<\sum_{t=1}^T\frac{\langle F_t \rangle}{(1+k)^t}.
	\end{equation}
\end{enumerate} 

\subsection{RADR: explanation}\label{RADRE}

Let us provide the explanation of the idea underlying this valuation method in terms of approach presented in the previous paragraphs. Let us assume that there exists a possibility of investment with some rate $k$. In such a case to obtain the cash flow $F_t$ in the period    $t$ one should make an initial outlay $I_I^{(t)}$ determined by the following expression:
\begin{equation}
I_I^{(t)}(1+k)^t=F_t.
\end{equation}  
The total initial outlay $I_I$ guaranteeing the the cash flow stream $\mathbf{F}=(F_1,\dots,F_T)$ is therefore
\begin{equation}
I_I=\sum_{t=1}^TI_I^{(t)}\equiv
\sum_{t=1}^T\frac{F_t}{(1+k)^t}
\equiv
{\rm PV}({\bf F}|k)
\end{equation}

Within the RADR approach one assumes that there is a possibility to make investments with (risky) rate $k$ replicating the mean cash flows $\{\langle F_t\rangle\}$:
\begin{equation}
I_t (\langle F_t \rangle,k) \cdot (1+k)^t = \langle F_t \rangle  \;\; \Rightarrow \;\; I_t (\langle F_t \rangle,k) = \frac{\langle F_t \rangle}{(1+k)^t}.
\end{equation}
However, as the rate $k$ is risky, the income from these partial investments is not guaranteed. The risk-free income that can be obtained from the partial investment outlay $I_I^{(t)}(\langle F_t\rangle|k)$  is determined by the risk-free rate $r$:
\begin{equation}
I_I^{(t)}(\langle F_t\rangle|k) \; \overset{ \text{r}}{\longmapsto} \; \left( \frac{1+r}{1+k} \right)^t \langle F_t \rangle \equiv \alpha (r,k)^t \langle F_t \rangle.
\end{equation}
This is the so-called {\it certainty equivalent} of the mean cash flow $\langle F_t \rangle$.  Therefore, the mean cash flows $\langle F_t \rangle$ can be represented as a composition of riskless and risky contributions:
\begin{equation}
\langle F_t \rangle =
\underbrace{\alpha(r,k)^t \langle F_t \rangle}_{\text{riskless}} 
+ 
\underbrace{(1-\alpha(r,k)^t) \langle F_t \rangle}_{\text{risky}}.
\end{equation}  
The distinguishing feature of the certainty equivalent is in the fact that its riskless present value is equal to the risky present value of the underlying mean cash flow:
\begin{equation}
{\rm PV}(\langle F_t\rangle|k)=
{\rm PV}(\alpha(r,k)^t\langle F_t\rangle |r)
\end{equation}
Thus, the acceptance criterion in the RADR formalism as expressed in terms of the certainty equivalents is expressed as follows:
\begin{equation}
-I_0+\sum_{t=1}^T{\rm PV}(\alpha(r,k)^t \langle F_t\rangle|r)>0,
\end{equation}  
or, equivalently,
\begin{equation}\label{crit_RADR}
-I_0+\sum_{t=1}^T{\rm PV}(\langle F_t\rangle|r)>
\sum_{t=1}^T{\rm PV}((1-\alpha(r,k)^t)\langle F_t\rangle|r).
\end{equation}
This means that the risk premium from the project should exceedthe outlay of the riskless investment guaranteeing risky part of the cash flows stream.  Let us note that
\begin{equation}
\langle{\rm NPV}(\mathbf{F}|r)\rangle \equiv
-I_0+\sum_{t=1}^T{\rm PV}(\langle F_t\rangle|r)
\end{equation}
and introduce the following notation
\begin{equation}
\Lambda^{\rm RADR}=\sum_{t=1}^T{\rm PV}((1-\alpha(r,k)^t)\langle F_t\rangle|r).
\end{equation}
In such a case, the criterion~\eqref{crit_RADR} takes the form
\begin{equation}
\langle {\rm NPV}({\bf F}|r)\rangle > \Lambda^{\rm RADR},
\end{equation}
which is equivalent to
\begin{equation}
\Omega( \Lambda^{\rm RADR})>1.
\end{equation}

Thus, the RADR evaluation imposes the exact value for the risk/profit separation scale $\Lambda=\Lambda^{\rm RADR}$. 

\subsection{RADR: discussion}\label{RADRD}

Both $\Lambda^{\rm RADR}$ and ${\rm NPV}^*$ (introduced in the paragraph~\ref{critval}) have a natural interpretation of critical values for ${\rm NPN}(\mathbf{F}|r)$, however there are a principal differences in their meanings.

\begin{itemize}
	\item The investment acceptance criterion 
	$$NPV(\mathbf{F}|r)>NPV^*$$ 
	refers to each trajectory separately, whereas the RADR criterion 
	$$\langle NPV(\mathbf{F}|r)\rangle>\Lambda^{\rm RADR}$$ 
	 operates with mean values only.
	
	\item The criterion $NPV(\mathbf{F}|r)>NPV^*$ is equivalent to the investment acceptance criterion in the form $\mu>\mu^*$. 
	
	In the case RADR approach one can write out the following chain of equivalent expressions\footnote{Where, in general, $(1+{\rm MIRR}(\mathbf{F}|k,k))^T=\frac{\sum_{t=1}^T F^+_t(1+k)^t}{I_0+\sum_{t=1}^T\frac{F^-_t}{(1+k)^t}}$. If only one investment cash flow $I_0$ exists $(1+{\rm MIRR}(\mathbf{F}|k))^T=\frac{\sum_{t=1}^T F^+_t(1+k)^t}{I_0}$}:
	\begin{equation}
	{\rm NPV}(\langle\mathbf{F}\rangle|k)>0
	\Longleftrightarrow
	{\rm MIRR}(\langle\mathbf{F}\rangle|k)>k
	\Longleftrightarrow
	\langle{\rm NPV}(\mathbf{F}|r)\rangle>\Lambda^{\rm RADR}
	\end{equation} 
	
	\item From the above equivalence it immediately follows that the RADR criterion implicitly fixes the reinvestment rate $k$, albeit for the mean cash flows only.
	
	\item In case of RADR there is only one choice for the minimal acceptable $\Omega$ level: $\Omega(\Lambda^{\rm RADR})>1$, as the comparison is made in terms of means. In general, $\Omega({\rm NPV}^*)>\Omega^*$, where $\Omega^*$ is equal to one only for the risk neutral investor and is less or greater than one for risk averse and risk seeking investors respectively.
	
	\item The approach presented in the previous section is also capable of accounting for sensitivity of the risk measure $\Omega$ to small changes in the value of the critical threshold (e.g. ${\rm NPV}^*$) which can be very useful for project's ranking. Such an analysis allows to choose the project with a "more stable/less sensitive"\; $\Omega$-ratio thus corresponding to a better risk profile. In contrast, in the RADR approach the threshold  $\Lambda^{\rm RADR}$  is fixed and there is no possibility to take into account the exact risk-profile of the project. 	 
\end{itemize}

	\section{Illustrative example}\label{example}
	
\begin{figure}[H]
	\centering
	\includegraphics[width=\linewidth]{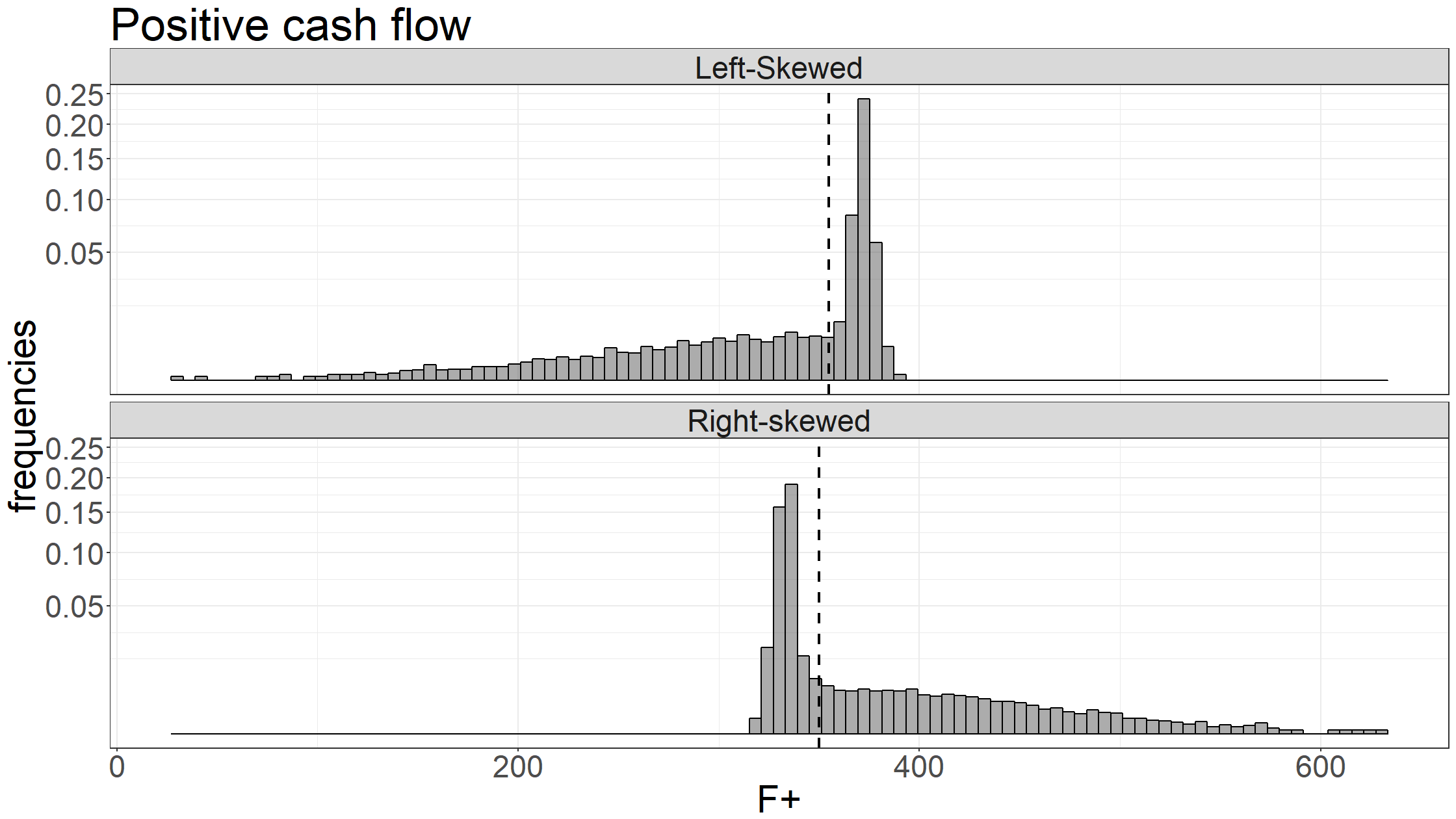}
	\caption{Distributions of $F^+$ for the two projects. Dashed vertical line shows the position of mean values}
	\label{fig:f+}
\end{figure}

	Let us consider two investment projects with the structure of cash flows shown in Table~\ref{tab:cashflow}, where the cash flow $F^+$ at time 1  is a random quantity while negative cash flows at times 0 and 2 are fixed. The projects differ in their distributions of $F^+$, the right-skewed for the first one, see Fig.~\ref{fig:f+}(Right-Skewed) and left-skewed for the second, see Fig.~\ref{fig:f+}(Left-Skewed). The characteristics of these distributions are shown in Table~\ref{tab:F+}. Let us note that the second "Left-Skewed" project has larger mean and median values.

		\begin{table}[H]
		\centering
		\begin{spacing}{1.5}		
			\begin{tabular}{|c|c|c|c|}
				\hline
				Time & 0 &1&2\\\hline
				Cash Flow & -200 & $F^+$ &-100\\\hline
			\end{tabular}
	\end{spacing}
	\caption{Cash flow structure}
	\label{tab:cashflow}
	\end{table}

\begin{table}[H]
	\centering
	\begin{spacing}{1.5}
	\begin{tabular}{|c|c|c|}
		\hline
		& Right-Skewed & Left-Skewed\\\hline
		Mean&350&355\\\hline
		Median&334&370\\\hline
		Std. Dev.&40&40\\\hline
		Skewness&2.7&-2.8\\\hline 
	\end{tabular}
	\end{spacing}
\caption{Characteristics of $F^+$ distributions for the two projects.}
\label{tab:F+}
\end{table}

Let us assume for simplicity the flat riskless discounting and forwarding rates of 5\% and that both projects are correlated with the market with the correlation coefficient $\rho$.
		
	\begin{enumerate}
		\item \textbf{RADR evaluation}
		
As standard deviations of cash flows in the two projects are the same, according to CAPM their $\beta$ coefficients are also equal ($\beta=\frac{\rho\sigma}{\sigma_m}$), where
$\sigma$ 	is a standard deviation of the projects return and $\sigma_m$  - that of the return of the market portfolio. Thus within the RADR framework the discounting rate for the two projects is the same, $$r^{RADR}=r_f+\beta(r_m-r_f),$$ where $r_f$- is the risk-free rate and $r_m$ - the return of the market portfolio. We have
$$
			{\rm NPV}(\langle{\bf F}\rangle\vert r^{RADR})=-200+\frac{\langle{F^+}\rangle}{1+r^{RADR}}-\frac{100}{(1+r^{RADR})^2},
$$ 		
where $\langle{F^+}\rangle$ is the average of $F^+$ (the value is provided in Table~\ref{tab:F+}).
			
Let us note that in the RADR/CAPM approach the "Left-Skewed" is better than the "Right-Skewed" one \textbf{for any} $\beta$ simply because of the ranking of the average values of positive cash flow. 
									
In Table~\ref{tab:RADR_crit} we compare characteritistics of both projects. We see that with $r^{RADR}=15\%$ ($\beta(r_m-r_f)=10\%$) the "Left-Skewed" project is always better than the "Right-Skewed" one. 	
							
			\begin{table}[h]
				\centering
				\begin{spacing}{1.5}
					\begin{tabular}{|c|c|c|}
						\hline
						Criterion & Right-Skewed & Left-Skewed\\\hline
						${\rm NPV}(\langle{\bf F}\rangle\vert 15\%)$&29&33\\\hline
						${\rm MIRR}(\langle{\bf F}\rangle\vert 15\%,15\%)$&20.8\%&21.7\%\\\hline
					\end{tabular}
			\end{spacing}
				\caption{Comparison of the two projects in the RADR/CAPM approach. RADR is equal to 15\%. ${\rm MIRR(\langle{\bf F}\rangle\vert 15\%,15\%)}$ is the MIRR value accounted with reinvestment and discount rates equal to $r^{RADR}=15\%$}
				\label{tab:RADR_crit}
			\end{table}
			
		\item \textbf{Evaluation in the new approach}
		
		Let investor's preferences be characterized by the desired risk premium of  $\Delta=10\%$ (e.g. $\Delta=\beta(r_m-r_f)$), i.e. critical $\mu$ equal to $\mu^*=15\%$. As negative cash flows and bond rate are fixed one can reconstruct the critical value of ${\rm NPV}$ (${\rm NPV}^*$). The corresponding values are shown in Table~\ref{tab:crit_val}.
		
		The histograms of the ${\rm NPV}({\bf F}\vert {\bf r})$ distributions of the projects are shown in Fig.\ref{fig:NPV}, where the vertical line shows the hurdle scale ${\rm NPV}^*$. The characteristics of the NPV distribution are shown in Table~\ref{tab:NPV}.	The distributions of $\mu$ are shown in Fig.~\ref{fig:MU} and the corresponding distribution characteristics in Table~\ref{tab:MU}.

		\begin{table}[H]
			\centering
			\begin{spacing}{1.5}
				\begin{tabular}{|c|c|}
				\hline
				Criterion & hurdle scale \\\hline
				$\mu^*$ & 15\% \\\hline
				${\rm NPV}^*$& 58 \\\hline
			\end{tabular}
			\end{spacing}
			\caption{Critical values of different characteristcis}
			\label{tab:crit_val}
		\end{table}	
		
		\begin{figure}[H]
			\centering
			\includegraphics[width=\linewidth]{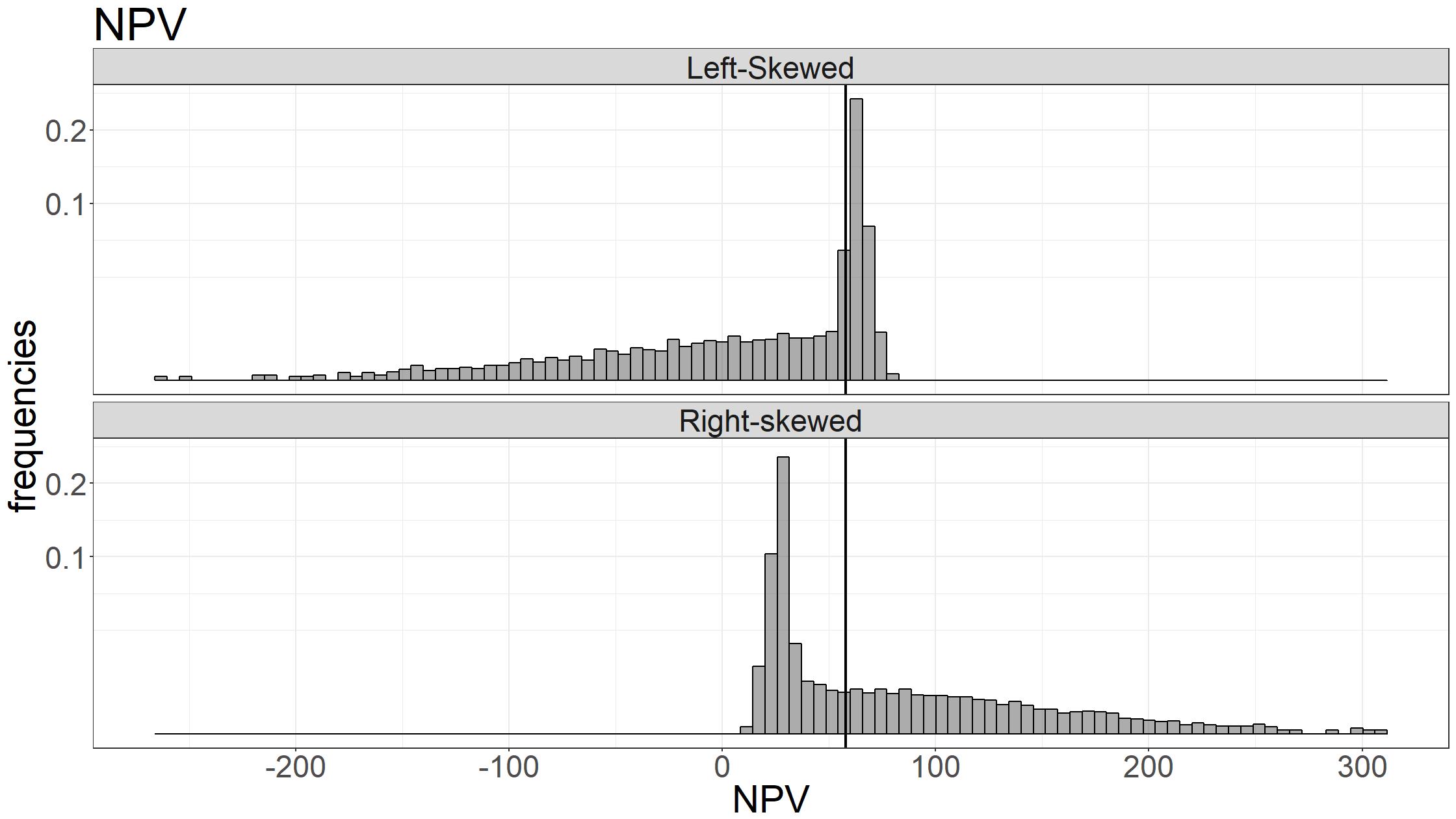}
			\caption{Distributions of ${\rm NPV}({\bf F}\vert{\bf r})$ in the two projects. Vertical line shows the position of ${\rm NPV}^*$}
			\label{fig:NPV}
		\end{figure}
	
		\begin{figure}[H]
		\centering
		\includegraphics[width=\linewidth]{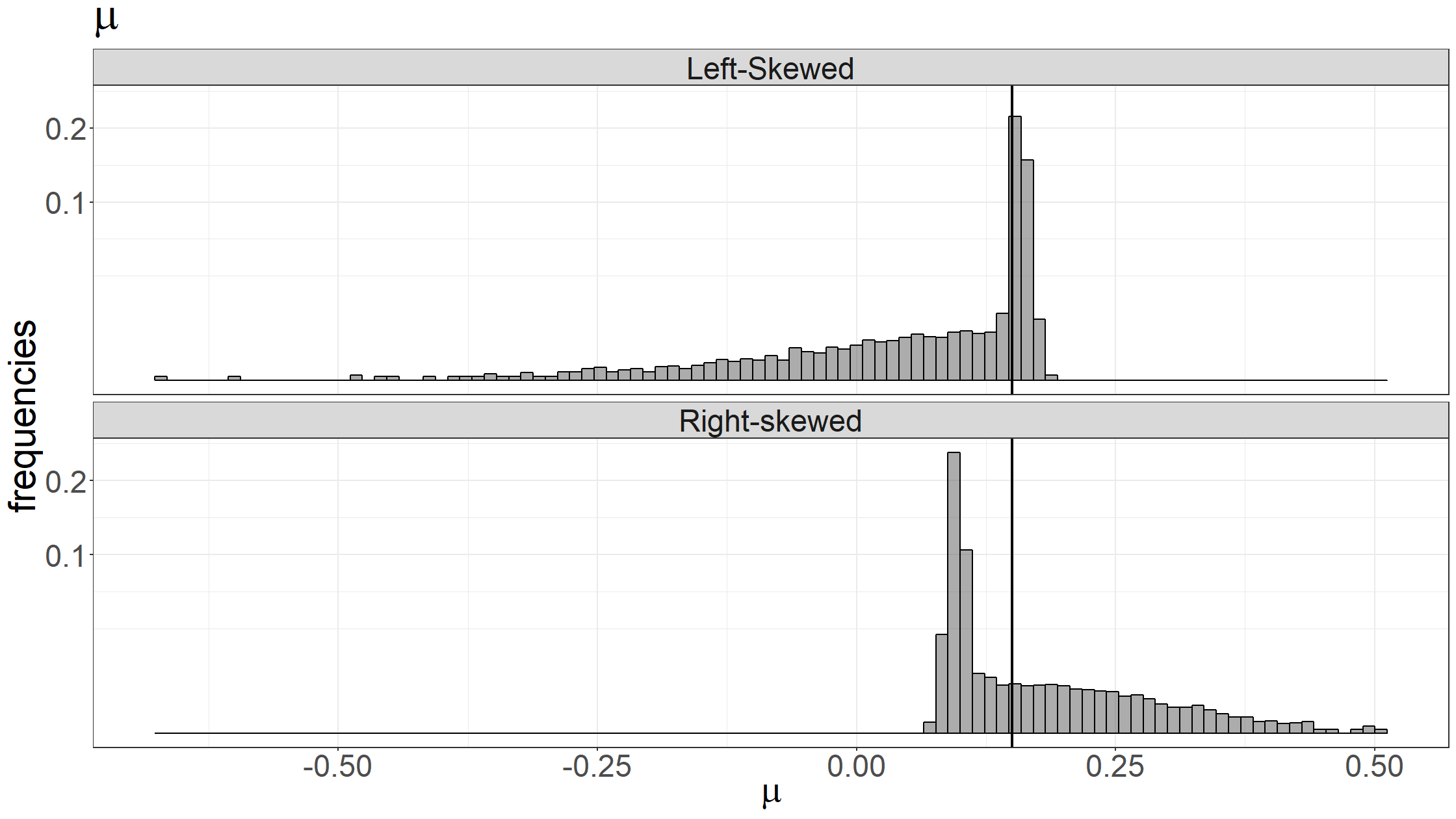}
		\caption{Distributions of $\mu$ in the two projects.  Vertical line shows the position of $\mu^*$}
		\label{fig:MU}
		\end{figure}
		
		\begin{table}[H]
			\centering
			\begin{spacing}{1.5}
				\begin{tabular}{|c|c|c|}
				\hline
				& Right-Skewed & Left-Skewed\\\hline
				Mean&42&47\\\hline
				Median&27&62\\\hline
				Std. Dev.&38&37\\\hline
				Skewness&2.7&-2.8\\\hline 
			\end{tabular}
			\end{spacing}
			\caption{Characteristics of the  ${\rm NPV}$ distributions for the two projects}
			\label{tab:NPV}
		\end{table}
	
		\begin{table}[H]
			\centering
			\begin{spacing}{1.5}
			\begin{tabular}{|c|c|c|}
			\hline
			& Right-Skewed & Left-Skewed\\\hline
			Mean&12.3\%&13.0\%\\\hline
			Median&9.9\%&15.7\%\\\hline
			Std. Dev.&0.06&0.07\\\hline
			Skewness&2.6&-3.2\\\hline 
		\end{tabular}
		\end{spacing}
		\caption{Characteristics of the $\mu$ distributions for the two projects}
		\label{tab:MU}
		\end{table}
		
		Knowing the critical values of ${\rm NPV}^*$ ($\mu^*$) one can calculate the value of $\Omega^{\rm NPV}$ ($\Omega^{\mu}$) for the two projects under consideration. The corresponding values are shown in Table~\ref{tab:omega}.

		\begin{table}[H]
			\centering
			\begin{spacing}{1.5}
				\begin{tabular}{|c|c|c|}
				\hline
				Project & $\Omega^{\rm NPV}$&$\Omega^{\mu}$ \\\hline
				Right-Skewed &  0.4&25.7\\\hline
				Left-Skewed &  0.3&3.1\\\hline
			\end{tabular}
			\end{spacing}
			\caption{The values of $\Omega$ for the two projects}
			\label{tab:omega}
		\end{table}	
		Therefore, with $\Delta$ of 10\% one should prefer the Right-Skewed project.
	\end{enumerate}

\section{Example of ranking of real industrial projects }
\label{real_ex}

Let us consider an example of ranking two real industrial projects  related to production of chemical fertilisers. Both projects were described in the form of excel table calculating project characteristics (e.g. project's NPV or $\mu$) from some inputs (e.g. price and macroeconomic indicators dynamic forecasts, plants characteristics, transportation tariffs forecasts etc.).   

\begin{figure}[H]
	\centering
	\includegraphics[width=\linewidth]{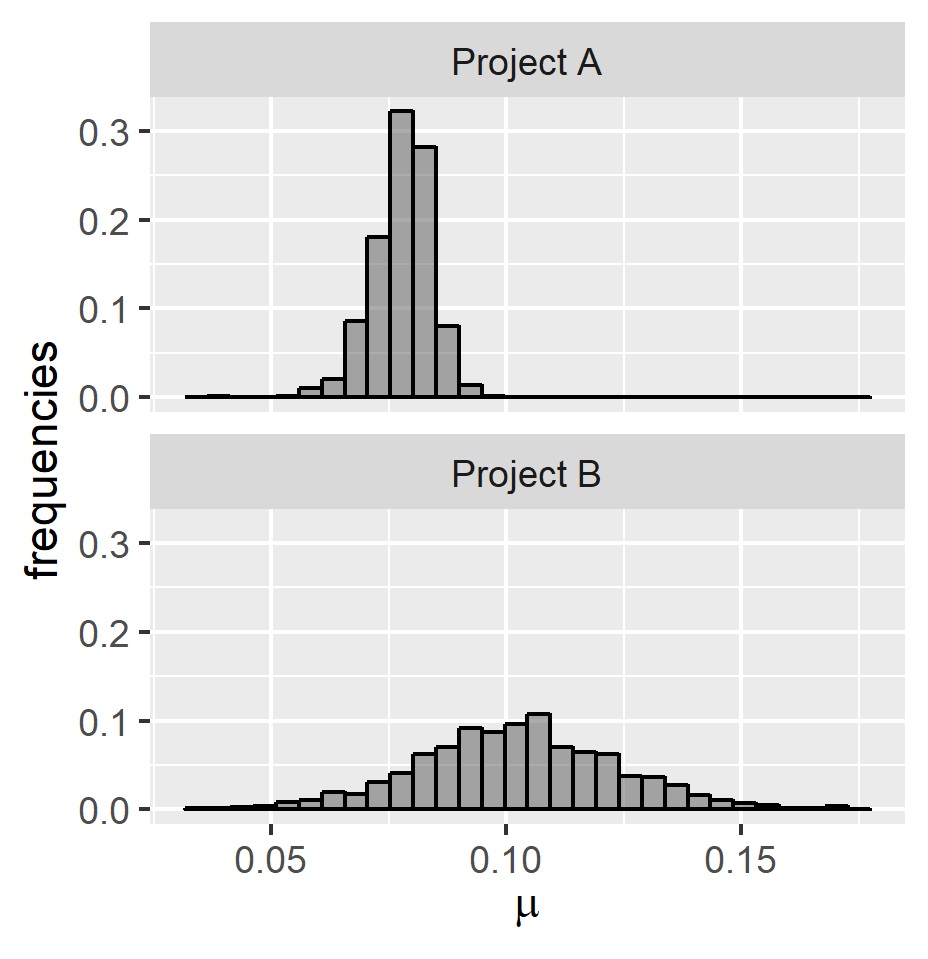}
	\caption{Distributions of $\mu$ of two real projects}
	\label{real_projects_hist}
\end{figure}

At the first stage we simulate 1000 Monte-Carlo scenarios for the models' inputs. After that we calculate different projects' characteristics, such as NPV and $\mu$, in each Monte-Carlo scenario. As a result of these procedure distributions of the projects under consideration were obtained. Histograms of these distributions are shown in Fig.~\ref{real_projects_hist}.

As it was described in the previous sections, the procedure of projects ranking consists of  three steps. The first is  to specify hurdle rate, the second is to calculate $\Omega$ using the chosen value and the third is to rank projects in decreasing order in $\Omega$.

\begin{figure}[H]
	\centering
	\includegraphics[width=\linewidth]{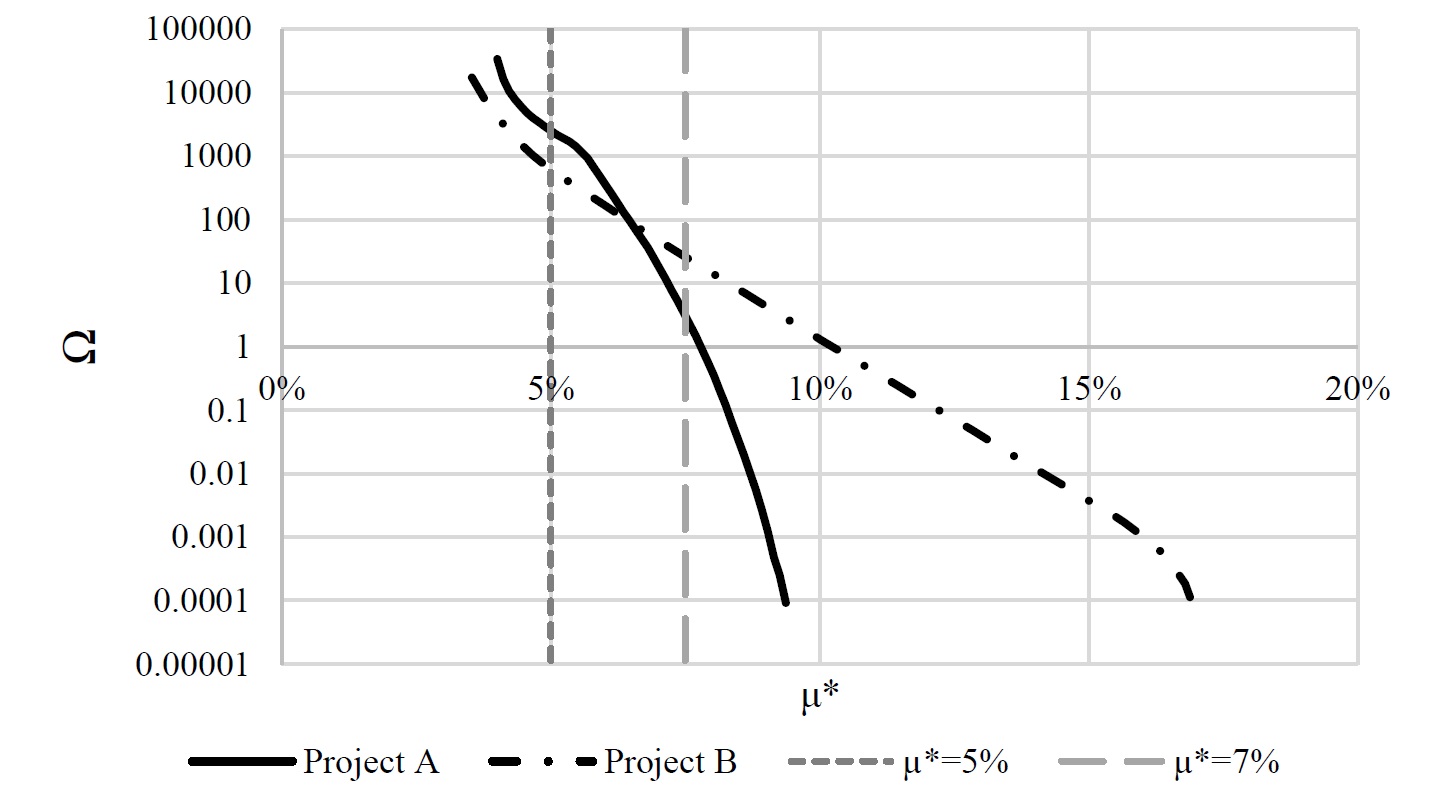}
	\caption{Distributions of $\mu$ of two real projects}
	\label{real_projects_comp}
\end{figure}

In Fig.~\ref{real_projects_comp} we show how values of $\Omega$ change with hurdle rate $\mu^*$. From this plot it follows that investors with different hurdle rates $\mu^*$  may have different projects ranking. For example, with $\mu^*=5\%$ the Project A is preferable, however, with $\mu^*=7\%$ an investor would choose the Project B.

\section{Conclusion}\label{conclusion}

The present study addresses one of the most important and, at the same time,  controversial problems in corporate finance -- evaluation and ranking of investment projects. The current industry standard is based on using for this purpose the risk-adjusted discount rate (RADR). It is however well known for quite a long time that the RADR methodology is plagued with serious limitations, e.g.:
\begin{itemize}
	\item it assumes some specific investor preferences that might not reflect the real ones;
	\item all the information on risks, i.e. on probabilistic description of premium, its moments, nature of its tail, etc. is compressed into one number, the risk premium, with no clear methodology of translating project-specific risks into this number,
\end{itemize}
etc. This absence of clear-cut methodology makes it very difficult to compare projects with different timespan, from different industries, etc. An investment certainty equivalence approach proposed in the present paper allows to perform en explicit separation of risky and riskless contributions to each possible realization of cash flows characterizing each particular investment project thus making it possible to apply modern criteria of evaluating and ranking of investment projects based on the corresponding exact distribution of risk premium. Detailed properties of these distribution are determined by such risk factors as sovereign, industry-specific or project-specific ones.

The approach makes it possible to 
\begin{itemize}
	\item compare investment projects from different industries through an assessment of differences in variability patterns of historical premiums of projects in these industries;
	\item use exact accounting for different risk sources resulting  in fully rational risk-adjustment selection;
	\item describe investor's risk-return preferences using only one parameter -- the hurdle rate, i.e. the premium scale that {\it for a given investor}  marks the range of  acceptable/non-acceptable risk premiums.  
\end{itemize}

The proposed methodology can be very helpful in organising a systematic procedure of evaluation and ranking of investment projects in large firms in which hundreds of investment projects with widely different timespans, economic significance and risk profiles are simultaneously considered.

Let us stress once again that in the present paper we discuss only the case of the riskless alternative investment. There remains a very important question of how to account for reinvestment risks. We plan to return to this question in future.

\medskip
	
\begin{center}
{\bf Acknowledgements}
\end{center}
We are very grateful to  A. Landia, V. Kalensky, A. Botkin, A. Djotyan and I. Nikola   for numerous discussions that were crucial for shaping our understanding of the subject of the present paper. 

\bibliography{Investment_riskless}
\bibliographystyle{APA}

		\appendix
		\section*{Appendix}
		
Let us define the minimally acceptable terminal  rate $\mu^*=r_T+\Delta^*_\mu$. Let us show, how to get the corresponding critical values for NPV. 	
From
\begin{eqnarray}
	(1+R^f_{T-t}) & = & \frac{(1+r_T)^T}{(1+r_t)^t},\\
	B_t(1+r_t)^t & = & F^+_t,
\end{eqnarray}
we get
\begin{eqnarray}
\label{eq:MIRR}
(1+\mu)^T=\frac{\sum_{t=0}^TF^+_t(1+R^f_{T-t})}{ I_0+\sum_{t=1}^{T}I_0^{(t)}}=
	\frac{\sum_{t=0}^{T}B_t(1+r_t)^t(1+R^f_{T-t})}{I_0+\sum_{t=1}^{T}I_0^{(t)} }=
	\\
	\frac{\sum_{t=1}^{T}B_t(1+r_t)^t\frac{(1+r_T)^T}{(1+r_t)^t}}{I_0+\sum_{t=1}^{T}I_0^{(t)} }=
	\frac{(1+r_T)^T\sum_{t=1}^{T}B_t}{I_0+\sum_{t=1}^{T}I_0^{(t)} }.
\end{eqnarray}
In addition, we have $\sum_{t=1}^{T}B_t={\rm NPV}(\mathbf{F}|\mathbf{r})+I_0+\sum_{t=1}^{T}I_0^{(t)}$ and, therefore,
\begin{equation}\label{mu_NPV}
	(1+\mu)^T=(1+r_T)^T\left(\frac{{\rm NPV}(\mathbf{F}|\mathbf{r})}{I_0+\sum_{t=1}^{T}I_0^{(t)}}+1\right)
\end{equation}
From  ~\eqref{mu_NPV} we finally get the relation between ${\rm NPV}^*$ and $\mu^*$ (and, therefore, $\Delta^*_{\mu}$)
\begin{equation}
		(1+\mu^*)^T=(1+r_T)^T\left(\frac{{\rm NPV}^*}{I_0+\sum_{t=1}^{T} I_0^{(t)}}+1\right),
\end{equation} 
i.e.
\begin{equation}
	{\rm NPV}^*=\left(\frac{(1+r_T+\Delta^*_{\mu})^T}{(1+r_T)^T}-1\right)\left(I_0+\sum_{t=1}^{T} I_0^{(t)}\right).
\end{equation}

\end{document}